\documentclass[final]{IEEEtran}
\usepackage{graphicx}
\usepackage[caption=false]{subfig}
\usepackage{cite}
\usepackage{color}

\usepackage{bm}
\usepackage{amssymb,amsmath}
\usepackage{verbatim}
\usepackage{acronym}
\acrodef{MIMO}{multiple input multiple output}
\acrodef{OFDM}{orthogonal frequency division multiplexing}
\acrodef{pdf}{probability density function}
\acrodef{KL}{Kullback-Leibler}
\acrodef{iid}{independent identically distributed}

\usepackage{tikz}
\usepackage[tikz]{mypics}

\newcommand\fig[1]{Fig.~\ref{f:#1}}
\newcommand\sfig[2]{Fig.~\ref{f:#1}\subref{f:#1#2}}

\newcommand\pmd{\beta}
\newcommand\pfa{\alpha}
\newcommand\KLdiv[2]{\ensuremath{\M D\left(#1\,||\,#2\right)}}
\newcommand\regU{\ensuremath{\C R^0}}

\newcommand\regq[1]{\ensuremath{\C R_{#1}}}
\newcommand\regOfD[1]{\ensuremath{\C R\left({#1}\right)}}
\newcommand\regint{\regq{\cap}}
\newcommand\opt{^\star}
\newcommand\trace{\mathrm{trace}\,}
\newcommand\bea[1][rCl]{\begin{IEEEeqnarray}{#1}}
\newcommand\eea{\end{IEEEeqnarray}}
\newcommand\sub[1]{_{\rm #1}}
\newcommand\super[1]{^{\rm #1}}
\newcounter{tempequationcounter}
\newtheorem{theorem}{Theorem}

\newtheorem{lemma}{Lemma}
\newtheorem{corollary}{Corollary}
\newtheorem{problem}{Problem}

\definecolor{ForestGreen}{rgb}{0.13, 0.55, 0.13}

\def\C{\mathcal}
\def\M{\mathcal}
\def\be{\begin{equation}}
\def\ee{\end{equation}}
\newcommand\e[1]{\label{#1}}
\newcommand\setg[1]{\left\{ #1 \right\}}
\def\tc{\,:\,}

\renewcommand\P[1]{{\rm P}[#1]}

\usepackage{revtools}
\showoldfalse
\shownotafalse
\definecolor{addcol}{named}{black}
\definecolor{addcolhl}{named}{white}

\definecolor{Royalblue}{cmyk}{1,0.30,0.2,0.2}

\title{On the Achievable Error Region of \\ Physical Layer Authentication Techniques \\ Over Rayleigh Fading Channels}

\author{Augusto Ferrante$^*$, Nicola Laurenti$^*$, Chiara Masiero$^*$,\\ Michele Pavon$^\dag$, and Stefano Tomasin$^*$\\
\normalsize $^*$ Department of Information Engineering, University of Padova, Italy \\
\normalsize $^\dag$ Department of Mathematics, University of Padova, Italy \\
Email: \{first\_name.last\_name\}@dei.unipd.it 
} 

\begin{document}	
\maketitle
\sloppy

\begin{abstract}
For a physical layer message authentication procedure based on the comparison of channel estimates obtained {from} the received messages, we focus on an outer bound on the type I/II error probability region. 
Channel estimates are modelled as multivariate Gaussian vectors, and we assume that the attacker has only some side information on the channel estimate, which {he} does not know directly. We derive the {attacking} strategy that provides the tightest bound on the error region, given the statistics of the side information. This turns out to be a zero mean, circularly symmetric Gaussian density whose correlation matrices {may be obtained by solving a constrained optimization problem}. We propose an iterative algorithm for its solution: Starting from the closed form solution of a relaxed problem, we obtain, by projection, an initial feasible solution; then, by an iterative procedure, we look for the fixed point solution of the problem. Numerical results show that for cases of interest the iterative approach converges, and perturbation analysis shows that the found solution is a local minimum.
\end{abstract}

\begin{IEEEkeywords}
Authentication, Physical layer security, Rayleigh fading channels, Hypothesis testing 
\end{IEEEkeywords}

%

\section{Introduction}

Physical layer security provides an effective defense mechanism which is complementary to higher layer security techniques. Indeed, it has the potential of resisting the attacks based on computational capabilities that may be feasible in the near future, e.g., by quantum computing. Moreover, security implemented at the physical layer is usually based on information theoretic arguments \cite{Bloch-book}. It therefore entails analytically predictable performance irrespective of the attacker capabilities and has found recently application to widely used communication systems \cite{Renna, Tomasin-MIMO}. One of the most desirable mechanisms of physical layer security is the authentication of the message source. This key task can be conveniently recast into a hypothesis testing problem \cite{Maurer00,Lai09}, namely to decide between hypothesis $\mathcal H_0$ that the message was effectively sent by the legitimate source, and hypothesis $\mathcal H_1$ that it was forged by the attacker.

Physical layer authentication has been addressed by considering either device-specific non-ideal transmission parameters extracted from the received signal \cite{Daniels05}, or channel characteristics to identify the link between a specific source and the receiver \cite{Faria06,Xiao09,Baracca12}. In this paper we focus on the latter case, which finds application in many wideband wireless systems, where even small changes in the position of the transmitter have a significant impact on the channel. In particular, we consider the approach of \cite{Baracca12}, where the test is performed in two phases. In the first phase, the receiver gets an authenticated noisy estimate $x$ of the channel with respect to the legitimate transmitter. In the second phase, upon reception of a message, the receiver gets a new estimate $u$ of the channel and compares it with $x$. Then, he  must decide whether $u$ is an estimate of the legitimate channel or the channel forged by an eavesdropper. 

The performance of a binary hypothesis testing scheme is measured by the probability of type I (false alarm), and type II (missed detection) errors. Therefore, theoretical bounds on the achievable error probability region are of great importance to establish the effectiveness of practical schemes. 
For instance, \cite{Maurer00} considered the traditional authentication scenario in which the legitimate parties can make use of a shared cryptographic key that is kept perfectly secret to the attackers. There, an outer bound on the achievable error region was derived, that holds irrespectively of the decision rule implemented by the receiver.  
Then, by fixing the false alarm probability, the outer bound is turned into a lower bound on the missed detection probability.
An analogous approach was used in \cite{Cachin98} and \cite{Barni13} within the different contexts of steganography and fingerprinting, respectively.
Similarly, in \cite{Lai09}, such lower bound is paired with an asymptotic upper bound, and both are derived also in the case that the legitimate parties share correlated sequences, instead of an identical key.
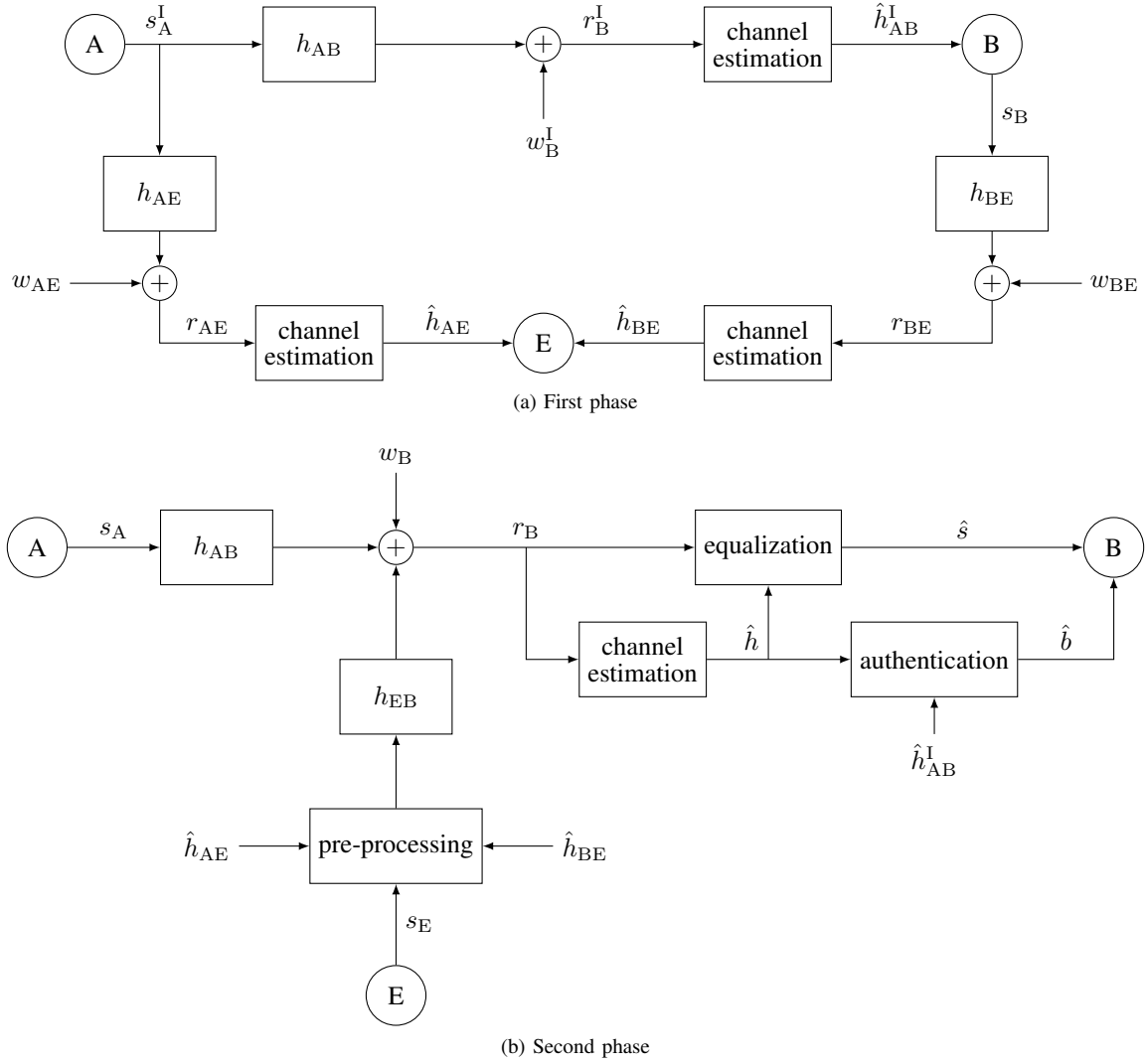
\begin{figure*}[t!]\newcommand\channel[2]{\node[rectangle,draw,minimum height = 10mm, minimum width = 15mm](h#1) at (#2){$h\sub{#1}$};}
\newcommand\process[3]{\node[rectangle,draw,minimum height = 10mm, minimum width = 15mm](#1) at (#3){#2};}
\newcommand\estim[2]{\node[rectangle,draw,minimum height = 10mm, minimum width = 15mm](est#1) at (#2){\shortstack{channel\\estimation}};}
\newcommand\source[2]{\node[circle,draw,minimum size = 8mm](#1) at (#2){#1};}
\newcommand\sumcir[2]{\node[circle,draw,inner sep=1pt,minimum size = 3mm](#1) at (#2){$+$};}
\begin{center}
\subfloat[First phase]{\label{f:chmodel1}%
\begin{tikzpicture}[x=6cm,y=2cm,>=latex]
\source{A}{0,2}
\source{B}{2,2}
\source{E}{1,0}
\channel{AB}{0.5,2}
\draw[->] (A) -- node[near start,above] (sA) {$s\sub A\super I$} (hAB);
\sumcir{wB}{1,2}
\draw[<-] (wB) --+ (0,-0.5) \wh t{w\sub B\super I};
\draw[->] (hAB) -- (wB);
\estim{B}{1.5,2}
\draw[->] (wB) -- node[near start,above](rB){$r\sub B\super I$} (estB);
\draw[->] (estB) -- node[midway,above](hhat) {$\hat h\sub{AB}\super I$} (B);
\channel{AE}{sA |- 0,1}
\draw[->] (sA) --  (hAE);
\sumcir{wAE}{hAE |- 0,0.4}
\draw[<-] (wAE) --+ (-0.2,0) \wh r{w\sub{AE}};
\draw[->] (hAE) --  (wAE);
\estim{AE}{hAB |- E}
\estim{BE}{estB |- E}
\draw[->] (wAE) |- node[near end, above](rAE){$r\sub{AE}$} (estAE);
\draw[<-] (E) -- node[midway,above](hhAE){$\hat h\sub{AE}$} (estAE);
\channel{BE}{B |- hAE}
\draw[->] (B) --  node[midway,right](sB){$s\sub B$} (hBE);
\sumcir{wBE}{hBE |- wAE}
\draw[<-] (wBE) --+ (0.2,0) \wh l{w\sub{BE}};
\draw[->] (hBE) --  (wBE);
\draw[->] (wBE) |- node[near end, above](rBE){$r\sub{BE}$} (estBE);
\draw[<-] (E) -- node[midway,above](hhBE){$\hat h\sub{BE}$} (estBE);
\end{tikzpicture}} \\
\subfloat[Second phase]{\label{f:chmodel2}%
\begin{tikzpicture}[x=6cm,y=2cm,>=latex]
\source{A}{0.2,2}
\source{B}{2.6,2}
\source{E}{1,-1}
\channel{AB}{0.6,2}
\draw[->] (A) -- node[midway,above] (sA) {$s\sub A$} (hAB);
\sumcir{wB}{1,2}
\draw[<-] (wB) --+ (0,0.5) \wh b{w\sub B};
\draw[->] (hAB) -- (wB);
\estim{B}{1.55,1.25}
\process{auth}{authentication}{2.2,1.25}
\process{equ}{equalization}{1.83,2}
\draw[<-] (auth) -- +(0,-0.5) node[below](hhAB){$\hat h\sub{AB}\super I$};
\draw[->] (wB) -- node[pos=0.4,above](rB){$r\sub B$} (equ);
\draw[->] (equ) -- node[midway,above](shat){$\hat s$} (B);
\draw[->] (rB) |- (estB);
\draw[->] (estB) -- (auth) node[above left](hhat) at (estB -| equ) {$\hat h$};
\draw[->] (auth) -| node[near start,above](bhat){$\hat b$}  (B);
\draw[->] (estB -| equ) -- (equ);
\process{g}{pre-processing}{E |- 0,0}
\draw[<-] (g) -- +(0.35,0) node[right](hhBE){$\hat h\sub{BE}$};
\draw[<-] (g) -- +(-0.35,0) node[left](hhAE){$\hat h\sub{AE}$};
\channel{EB}{wB |- hBE}
\draw[->] (E) -- node[midway,right](sE){$s\sub E$}(g);
\draw[->] (g) -- (hEB);
\draw[->] (hEB) -- (wB);
\end{tikzpicture}}
\end{center}
\caption{Transmission channel scenario for the physical layer authentication problem.}\label{f:chmodel}\end{figure*}

In the above cases, since the attacker has no information on the shared sequences, the optimal attack strategy with respect to the outer bound is to present forged signals that, albeit independent of the legitimate shared key, are generated from the same marginal distribution as the legitimate signals.
In our framework, on the contrary, the legitimate authentication signal is the actual realization of a fading wireless channel. Thus the attacker has some side information given by the channel estimates $z$ he performs, which are in general correlated with the legitimate channel.
We model channel estimates as {correlated} multivariate Gaussian vectors, which is a usual assumption in wireless transmissions, including those using \ac{OFDM} or \ac{MIMO}.

{The contribution of our paper is thus threefold:
1) we derive an outer bound to the error probability region, in terms of the attacker strategy; {2) we prove the existence of a strategy $v$, jointly Gaussian with $z$, that yields the tightest bound, and characterize the covariance through the solution of a system of two matrix equations}; 3) we give an efficient technique for the numerical evaluation of the optimal attack strategy and the corresponding bound.} 

The paper is organized as follows. Section \ref{sec:prob_stat} introduces the problem formally, so that the theoretical results can be derived in Section \ref{sec:main_res}. Based on those results, in Section \ref{sec:algorithm} we propose an efficient algorithm for the numerical evaluation of the optimal attack strategy. Then, in Section \ref{sec:numerical_results} we give examples of numerical results, and eventually we draw conclusions in Section \ref{sec:conc}.

In our notation, if $a \in \M C^n$ and $b\in \M C^m$ are random vectors, $K_{ab}$ denotes the $n\times m$ covariance matrix of vectors $a$ and $b$, whereas $K_{\bigl[\begin{smallmatrix}a\\b\end{smallmatrix}\bigr]}$ stands for the $(n+m)\times (n+m)$ variance matrix of the vector~$\bigl[ \begin{smallmatrix}a\\b \end{smallmatrix}\bigr]$.
The symbol $A^\ast$ denotes the complex conjugate transpose of matrix $A$.


\section{Problem Statement}\label{sec:prob_stat}


We consider the physical layer channel authentication scheme depicted in \fig{chmodel} where agents Alice (A) and Eve (E) transmit messages to Bob (B), and Bob aims at authenticating messages from Alice, i.e., reliably detecting whether she sent them or not. The authentication is performed via a two phase procedure, as detailed in \cite{Baracca12}:
\paragraph{First Phase} In this phase, illustrated in \sfig{chmodel}{1} Alice transmits one or more messages, denoted by $s\sub A\super I$, whose authenticity is guaranteed by higher layer techniques, to Bob, through the channel $h\sub{AB}$. Bob gets a noisy estimate $\hat h\sub{AB}\super{I}$ of the channel with respect to Alice and replies with an acknowledgement message. Moreover, by leveraging transmissions by Alice and Bob, Eve obtains (possibly noisy) estimates $\hat h\sub {AE}, \hat h\sub{BE}$ {of} the channels that link her to both agents.
\paragraph{Second Phase} Subsequently, as shown in \sfig{chmodel}{2}, either Alice or Eve transmit messages $s\sub A$ or $s\sub E$, respectively. Bob authenticates the received messages by getting a new noisy channel estimate $\hat h$ and comparing it with his template $\hat h\sub{AB}\super{I}$. If this decision process $\C D$ deems the message as coming from Alice, the binary flag $\hat b$ is set to zero, otherwise it is set to one. In this phase, Alice performs transmissions in a similar fashion as the first phase, yet the new estimate $\hat h\sub{AB}\super{II}$ of the Alice-Bob channel will not be identical to $\hat h\sub{AB}\super{I}$, in general, due to the independent noises that affect both estimates. On the other hand, Eve can perform a pre-processing of her own messages in order to induce an equivalent channel estimate by Bob that is as close as possible to $\hat h\sub{AB}\super{I}$. 

\begin{figure}\newcommand\channel[3]{\node[rectangle,draw,minimum height = 10mm, minimum width = 15mm](#1) at (#3){$#2$};}
\newcommand\source[3]{\node[circle,draw,minimum size = 8mm](#1) at (#3){$#2$};}
\newcommand\sumcir[2]{\node[circle,draw,inner sep=1pt,minimum size = 3mm](#1) at (#2){$+$};}
\begin{center}
\begin{tikzpicture}[x=3cm,y=3cm,>=latex]
\source{pxyz}{p_{xyz}}{1,0}
\channel{pvz}{p_{v|z}}{0,-1}
\channel{D}{\C D}{2,-1}
\channel{Hb}{\C H_b}{1,-1}
\draw[->](pxyz) -| node[near start,above]{$x$} (D);
\draw[->](pxyz) -- node[midway,right]{$y$} (Hb);
\draw[->](pxyz) -| node[near start,above]{$z$} (pvz);
\draw[->](pvz) -- node[midway,above]{$v$} (Hb);
\draw[->](Hb) -- node[midway,above]{$u$} (D);
\draw[->](D) -- node[midway,above]{$\hat b$} +(0.6,0);
\end{tikzpicture}
\end{center}
\caption{Abstract model for physical layer authentication cast as an hypothesis testing problem with channel estimates as the available observations.}\label{f:sysmodel}\end{figure}
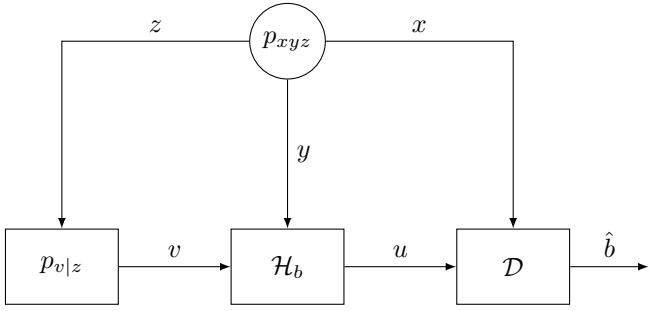
From now on, for the sake of a more compact notation, we let $x = \hat h\sub{AB}\super I$, $y = \hat h\sub{AB}\super{II}$, $z = (\hat h\sub{AE},\hat h\sub{BE})$, $u = \hat h$ and we refer to the abstract representation of the authentication scenario given in \fig{sysmodel}. There, the joint \ac{pdf} $p_{xyz}$ of the channel estimates is determined by the fading environment and the estimation techniques adopted by the agents, which are assumed known by all of them. In order to consider a worst case scenario, we assume that Eve is able to forge any equivalent channel estimate $v$ on Bob, neglecting the fact that power constraints and/or channel characteristics may prevent this and restrict the set of possible attacks, in practice. As a side effect, this assumptions also allows to simplify our analysis. The attacker's forging strategy can make use of the information carried by her observations $z$, and in order to allow her the most generality, we consider that she can make use of a probabilistic strategy, which is thus characterized by the conditional \ac{pdf} $p_{v|z}$. 
Note that, although our framework considers a single forging attempt, it can be extended to a sequence of attempts $\setg{v_i}$, $i=1, 2, \ldots$,  where the attacker strategy is represented by the family of conditional \acp{pdf} $\setg{p_{v_i|z,v_1,\ldots,v_{i-1}}}$. 

Given the channel estimate $u$, Bob decides between the two hypotheses
\bea[r'c'l"s]
\C H_0 & : & u = y & message is from Alice \label{e:hypothesis0}, \\
\C H_1 & : & u = v & message was forged. \label{e:hypothesis1}
\eea
In \fig{sysmodel}, being in hypothesis $\C H_0$ or $\C H_1$ is obtained by setting $b=0$ or $1$, respectively. Correct authentication is achieved when $\hat{b} = b$.

Recall that all channels are described by zero-mean circular symmetric complex Gaussian vectors with correlated entries, as a suitable model for many scenarios (including \ac{MIMO}/\ac{OFDM}). Moreover, we assume that all transmissions are corrupted by additive white Gaussian noise with zero mean. Similarly, we assume that also the channel estimates are zero-mean circular symmetric complex Gaussian vectors with correlated entries.\footnote{This is justified by the fact that, in order to be effective, estimates should have a distribution that is close to that of the target variable. Furthermore, under mild assumptions on the SNR and with a sufficient amount of data, errors in, e.g. an ML estimation, are asymptotically unbiased, efficient and Gaussian distributed themselves \cite[\S 7.8]{Kay93}.} In particular, {$x$, and $y$  are  $n$-dimensional, complex, circular symmetric Gaussian random vectors, $z$ is an  $m$-dimensional, complex, circular symmetric Gaussian random vector. On the other hand $v$ is an $n$-dimensional, complex, random vector whose probability density is not specified as it will be chosen by the attacker in order to obtain better mimetic features.}
We denote the set of all possible conditional distributions (forging strategies) $p_{v|z}(\cdot|\cdot)$ as
\be
\C Q = \setg{q(\cdot|\cdot) \,:\, \M C^n\times \M C^m \to \M R, q(b|c) \geq 0, \int q(b|c)\,{\rm d}b = 1}\,.
\ee

Performance of the authentication system are assessed by type I error probability $\pfa$, i.e., the probability that Bob discards a message as forged by Eve while it is coming from Alice
\begin{equation}
\pfa = \P{\hat b = 1|\C H_0}\,,
\end{equation}
and the type II error probability $\pmd$, i.e., the probability that Bob accepts a message coming from Eve as legitimate
\begin{equation}
\pmd = \P{\hat b = 0|\C H_1}\,.
\end{equation} 

The aim of a clever design for the authentication scheme is to make both error probabilities $\pfa$ and $\pmd$ as small as possible. Since it is trivial to achieve $\pfa + \pmd= 1$ with a random decision strategy that outputs  $\hat b = 1$ with probability $\pfa$, independently of the observation $u$, we are only interested in values of $\pfa$, $\pmd$ in the region
\begin{equation}
\regU = \setg{(\alpha,\beta) \tc \alpha\geq 0, \beta\geq 0, \alpha+\beta \leq 1 }\,.
\end{equation}

\subsection{Error Region Bounds for a Given {Attacking} Strategy}


A first bound on the error region for a given {attacking} strategy can be obtained by applying the fundamental data processing inequality for the \ac{KL} divergence \cite{Kullback67} to our binary hypothesis decision scheme $\C D$. In fact, from \cite{Cachin98,Maurer00} we have\footnote{Note that the symmetric bound $\KLdiv{p_{\hat b|\C H_0}}{p_{\hat b|\C H_1}} \leq  \KLdiv{p_{xu|\C H_0}}{p_{xu|\C H_1}}$ holds as well (see also \cite{Baracca12}).}
\be 
\begin{split}
\KLdiv{p_{\hat b|\C H_1}}{p_{\hat b|\C H_0}} \leq   \KLdiv{p_{xu|\C H_1}}{p_{xu|\C H_0}}\,.
\end{split}
\e{bound}
\ee 
First we observe that $p_{\hat b|\C H_0}(1) = \alpha$, $p_{\hat b|\C H_0}(0) = 1 - \alpha$, and similarly $p_{\hat b|\C H_1}(0) = \beta$, $p_{\hat b|\C H_1}(1) = 1 - \beta$. Therefore, introducing 
the function\footnote{Notice that $f(\varphi, \psi)$ is the KL divergence between two Bernoulli probability
distributions of parameters $\varphi$ and $1-\psi$, respectivley.} 
\begin{equation}
f\left(\varphi,\psi\right) = \varphi\log\frac{\varphi}{1-\psi} + (1-\varphi)\log\frac{1-\varphi}{\psi}\,, \quad \varphi,\psi  \in [0,1]
\label{defd}
\end{equation}
we can rewrite \eqref{bound} as
\be 
\begin{split}
f(\beta, \alpha) \leq   \KLdiv{p_{xu|\C H_1}}{p_{xu|\C H_0}}\,.
\end{split}
\e{bound3}
\ee

{Since the observation $z$ encloses all the information the attacker can exploit in order to deceive the receiver, we can assume that the forging strategy $v$ is \emph{conditional independent} of the secure template $x$, given $z$. 
Then the divergence on the right side of \eqref{bound3} can be written explicitly for a given {attacking} strategy $q(\cdot|\cdot) \in \C Q$ as}
\begin{equation}
\begin{split}
D(q) & =   \KLdiv{p_{xu|\C H_1}}{p_{xu|\C H_0}} = \KLdiv{p_{xv}}{p_{xy}} \\ 
& =  \int\!\!\int \left[\int p_{xz}(a,c) q(b|c)\,{\rm d}c\right] \times \\
& \!\left[\log \left(\int p_{xz}(a,c) q(b|c)\,{\rm d}c\right)- 
\log p_{xy}(a,b)\right] {\rm d}a\,{\rm d}b\,.
\end{split}
\e{KLq2}
\end{equation}
Let {$f_0\geq 0$ be  given and set}
\be\label{eq:regd2}
\regOfD{f_0} :=   \setg{(\alpha,\beta) \in \regU \tc f(\beta,\alpha) \leq f_0}.
\ee
 Then \eqref{bound3} can be rewritten as
\be
(\alpha,\beta) \in \regOfD{D(q)}\ .	\e{bound2}
\ee
%
%

\subsection{Error Region Bounds for Any {Attacking} Strategy}
Each outer bound in \eqref{bound2} is clearly looser than 
\bea 
\regint & = & 
\bigcap_{q\in\C Q}\regOfD{D(q)} = \regOfD{D\opt} \e{boundint}
\eea
where
\bea
D\opt &=& \inf_{q\in Q} D(q)\,. \e{inf2}
\eea
Note that the region in \eqref{boundint} is not strictly speaking an outer bound of the type \eqref{bound2}, since the infimum \eqref{inf2} may not be achievable, in general. In that case, \eqref{boundint} represents  
a worst case performance for the authentication system, over all possible attacking strategies. 
On the other hand, for the attacker, approaching \eqref{inf2} represents the possibility to effectively carry out an impersonation attack.

The main goal of this paper it to evaluate the tightest bound \eqref{boundint}. Indeed, we provide an {attacking} strategy achieving \eqref{inf2}, under the assumption that the observation $z$ encodes all the information about the secure template $x$ the attacker can rely on in order to deceive the receiver.
{We have just shown that this is equivalent to the following constrained optimization problem:
\begin{problem}\label{prob:preci-math-prob}
Given the zero-mean, circular symmetric,  jointly Gaussian random vectors $x,y,z$ with joint covariance matrix 
\begin{equation}\label{eq:K_xyz}
K_{\Bigl[\begin{smallmatrix}x\\y\\z\end{smallmatrix}\Bigr]} := \begin{bmatrix} K_{xx} & K_{xy} & K_{xz}\\ K_{yx} & K_{yy} & K_{yz} \\ K_{zx} & K_{zy} & K_{zz}  \end{bmatrix},
\end{equation}
find a joint probability distribution $p_{xvz}\in L^1(\mathbb{C}^{2n+m})$ such that its marginal $p_{xv}$ minimizes  $\KLdiv{p_{xv}}{p_{xy}}$ under the constraints:\\
1.  The marginal distribution of $x,z$ (corresponding to $p_{xvz}$)  is equal to the given distribution $p_{xz}$.\\
2. The random vectors $v$ and $x$ are conditionally independent given $z$.
\end{problem}
} 

\section{Main results}\label{sec:main_res}
{In this section, we address Problem \ref{prob:preci-math-prob}.
In particular, we show that the problem is feasible, that it admits an optimal solution and that this solution is Gaussian. Finally, we show how to reformulate this problem in terms of solutions of two coupled matrix equations.
The first issue to be considered is the {\em feasibility} of Problem \ref{prob:preci-math-prob}, namely the existence of a
distribution $p_{xvz}$ satisfying the constraints and such that $\KLdiv{p_{xv}}{p_{xy}}$ is finite.
\begin{lemma}\label{feasibility}
Problem \ref{prob:preci-math-prob} is feasible.
\end{lemma}
\begin{IEEEproof}
Let $v$ be an $n$-dimensional, complex, zero-mean, circular symmetric Gaussian random vector (with arbitrary covariance) independent of $x$ and of $z$.
It is immediate to check that the corresponding $p_{xvz}$ satisfies the constraints and is such that $\KLdiv{p_{xv}}{p_{xy}}$ is finite.
\end{IEEEproof}
}

\begin{lemma}\label{lem:cond_ort}
{Let $x$ and $z$ be jointly Gaussian.} For any {attacking} strategy {$p_{xv}$ having finite second moment and} in which $v$ and $x$ are conditionally independent given $z$, they are also conditionally orthogonal given $z$, that is  
\begin{equation}\label{eq:partial_incorrelation}
\mathbb{E} \left[(x-\bar{\mathbb{E}}[x|z])(v-\bar{\mathbb{E}}[v|z])\right] = 0,
\end{equation}
where $\bar{\mathbb{E}}[\cdot|z]$ stands for the best linear estimator of $\cdot$ given $z$
\end{lemma}
\begin{IEEEproof}
We have
\begin{align}
\mathbb{E} \left[(x-\bar{\mathbb{E}}[x|z])(v-\bar{\mathbb{E}}[v|z])\right]   \\
= \mathbb{E} \left[\mathbb{E} \left[(x-\bar{\mathbb{E}}[x|z])(v-\bar{\mathbb{E}}[v|z])|z\right]\right]\label{eq:TET}\\
= \mathbb{E} \left[\mathbb{E} \left[(x-\bar{\mathbb{E}}[x|z])|z\right]\mathbb{E} \left[(v-\bar{\mathbb{E}}[v|z])|z\right]\right] \label{eq:cond_ind}\\
 =\mathbb{E} \left[\left(\mathbb{E} [x|z]-\bar{\mathbb{E}}[x|z]\right)\left(\mathbb{E}[v|z]-\bar{\mathbb{E}}[v|z]\right)\right]\label{eq:joint_gaussian} ,
\end{align}
where \eqref{eq:TET} and \eqref{eq:cond_ind} follow from the Total Expectation Theorem and the definition of conditional independence, respectively.
Since $x$ and $z$ are jointly Gaussian, we have that $\mathbb{E} \left[x|z\right] = \bar{\mathbb{E}}[x|z]$. Thus, we can conclude that the right-hand side of \eqref{eq:joint_gaussian} is equal to zero.
\end{IEEEproof}
 
In general, conditional independence does not imply conditional orthogonality, although for jointly Gaussian variables they are equivalent. However, we have proved that conditional independence of $x$ and $v$ given $z$ implies that $x$ and $v$ are conditionally orthogonal given $z$, thanks to $x$ and $z$ being jointly Gaussian.  

{
Let us recall the joint covariance matrix (\ref{eq:K_xyz})
\begin{equation}\label{eq:K_xvz}
K_{\Bigl[\begin{smallmatrix}x\\v\\z\end{smallmatrix}\Bigr]} := \begin{bmatrix} K_{xx} & K_{xv} & K_{xz}\\ K_{vx} & K_{vv} & K_{vz} \\ K_{zx} & K_{zv} & K_{zz}  \end{bmatrix}.
\end{equation}
Notice that, since the attacker knows the joint probability density $p_{xyz}$, the corner elements of \eqref{eq:K_xvz} are known. For the sake of simplicity, we introduce the following symbols for the unknown blocks of \eqref{eq:K_xvz}:
\begin{gather}
X := K_{vv}, \quad Y:=K_{xv},\quad  Z:=K_{vz}.
\end{gather}
Hence, we can write 
\begin{equation}\label{eq:K_xvz_inc}
K_{\Bigl[\begin{smallmatrix}x\\v\\z\end{smallmatrix}\Bigr]}=\begin{bmatrix} K_{xx} & Y & K_{xz}\\ Y^* & X & Z \\ K_{xz}^* & Z^* & K_{zz}  \end{bmatrix}.
\end{equation}
Recall that the conditional orthogonality of $x$ and $v$ given $z$ is equivalent to the following zero-block pattern in its inverse\footnote{A proof can be worked out in the same vein of \cite[Section~2]{Speed86}.}
\begin{equation}
\label{eq:K_xvz_inv}
K_{\Bigl[\begin{smallmatrix}x\\v\\z\end{smallmatrix}\Bigr]}^{-1} = \begin{bmatrix} * & 0 & *\\ 0 & * & * \\ * & * & * \end{bmatrix}.
\end{equation}   
{In this way we have expressed the second constraint of Problem \ref{prob:preci-math-prob} in terms of the structure of the inverse of the covariance matrix. 
We can therefore enforce this constraint by resorting to a ``maximum entropy" completion as described in \cite{Dempster-72}, see also \cite{Ferrante-P-IEEE-IT-11} for a more general result and \cite{CFPP-IEEE11} for an application of this technique.}
}
{
\begin{lemma}\label{lem:opt_gauss}
If $q_{\rm G}$ is a circular symmetric Gaussian distribution, then, among all distributions $p$ that share 
the same mean vector $\mu$ and covariance matrix $K$,
the one that minimizes $\KLdiv{p}{q_{\rm G}}$ is circular symmetric and Gaussian. 
\end{lemma}
\begin{IEEEproof}
Let $p_G$ be a \emph{circular symmetric Gaussian} probability density on $\mathbb{C}^n$ and let $p\neq p_G$ be any density having the same first and second moment as $p_G$. We denote by $H(p)$ the differential entropy of the density $p$, i.e. $H(p):=-\int p(a)\log p(a)\,da$.
Then (see \cite[Theorem~2]{Neeser93}), we have the inequality
\begin{equation}\label{entropy inequality}
H(p)<H(p_G).
\end{equation}
Now let $q_G$ be any proper Gaussian density on $\mathbb{C}^n$. 
Under the same hypotheses, we have
\begin{equation}\label{equality}
\int \log q_G(x)p(x)dx=\int\log q_G(x)p_G(x)dx,
\end{equation}
because $\log q_G(x)$ is a quadratic function of $x$. 
In view of (\ref{entropy inequality}) and (\ref{equality}), we now have
\begin{equation*}
\begin{split}
\mathbb{D}(p\|q_G) &=\int \log\frac{p(x)}{q_G(x)}p(x)dx\\
                   &=-H(p)-\int\log q_G(x)p(x)dx\\
                   &=-H(p)-\int\log q_G(x)p_G(x)dx\\
                   &\geq-H(p_G)-\int\log q_G(x)p_G(x)=\mathbb{D}(p_G\|q_G),
\end{split}
\end{equation*}
with equality iff $p_G$ is circular symmetric and Gaussian.
Thus, if $p$ is the solution of any minimum entropy problem with circular symmetric Gaussian prior, $p$ has to be circular symmetric and Gaussian.
\end{IEEEproof}
}
{
\begin{lemma}\label{ottimofinito}
If the second moment of $p_{xv}$ is not finite then $\KLdiv{p_{xv}}{p_{xy}}=\infty$.
\end{lemma}
\begin{IEEEproof}
We assume that $\KLdiv{p_{xv}}{p_{xy}}$ is finite and show that the second moment of $p_{xv}$ is finite.
Let us first recall the variational formula for the relative entropy \cite[page 68]{Deuschel-Stroock}: 
\begin{equation}
\label{vfre}
\begin{split}
\KLdiv{p_{xv}}{p_{xy}}=\sup_{\varphi\in\Phi}
\left\{\int_{\mathbb{C}^{2n}}\varphi(a) p_{xv}(a) da - \right.\\ 
\left.\log\left[\int_{\mathbb{C}^{2n}}\exp[\varphi(a)]p_{xy}(a)da \right]\right\}
\end{split}
\end{equation}
where $\Phi$ is the set of bounded functions.
Observe now that, since $p_{xy}$ is a Gaussian probability density, there exists $\varepsilon>0$ such that
$$L:=\mathbb{E}_{p_{xy}}[\exp[\varepsilon \|a\|^2]]=\int_{\mathbb{C}^{2n}}
\exp[\varepsilon \|a\|^2] p_{xy}(a) da$$ 
is finite.
Let us now consider the following sequence of bounded functions: 
$$\varphi_l(a):=\left\{\begin{array}{ll}\varepsilon \|a\|^2,\ &{\rm if\ } \|a\|^2\leq l,\\
0,\ &{\rm if\ } \|a\|^2>l.\end{array} \right. $$ 
From (\ref{vfre}) we get that for all $l=1,2,\dots$,
\begin{equation}\label{finiteinequality0}
\begin{split}
\KLdiv{p_{xv}}{p_{xy}}+\log\left[\int_{\mathbb{C}^{2n}}\exp[\varphi_l(a)]p_{xy}(a)da \right]\geq  \\
\int_{\mathbb{C}^{2n}}\varphi_l(a) p_{xv}(a) da,
\end{split}
\end{equation}
or, equivalently,
\begin{equation}\label{finiteinequality}\begin{split}
\frac{1}{\varepsilon}\left\{\KLdiv{p_{xv}}{p_{xy}}+\log\left[\int_{\mathbb{C}^{2n}}\exp[\varphi_l(a)]p_{xy}(a)da \right]
\right\}\geq \\
\int_{\Omega_l}\|a\|^2 p_{xv}(a) da,
\end{split}\end{equation}
where $\Omega_l:=\{a\in\mathbb{C}^{2n}:\ \|a\|^2\leq l\}$. 
As $l\rightarrow\infty$, the left-hand side of (\ref{finiteinequality}) converges to  $\frac{1}{\varepsilon}[\KLdiv{p_{xv}}{p_{xy}}+L]$ while the right hand side converges to the trace of the second moment of $p_{xv}$.
Such a trace is therefore finite and thus also the second moment of $p_{xv}$  is finite.
\end{IEEEproof}
We are now ready to consider the {\em existence} problem. As in many optimization problems this is one of the most delicate issue.
\begin{theorem}\label{existence}
There exists an optimal solution $p\opt_{xv}$ of Problem \ref{prob:preci-math-prob}.
\end{theorem}
\begin{IEEEproof}
Let $d\opt$ be the infimum of $\KLdiv{p_{xv}}{p_{xy}}$ over $p_{xv}$, satisfying the constraints of Problem \ref{prob:preci-math-prob}.
Let $p^j_{xvz}$, $j=1,2,\dots$, be a sequence of probability densities satisfying the constraints of Problem \ref{prob:preci-math-prob} and such that the corresponding marginals $p^j_{xv}$ satisfy  
$$
\lim_{j\rightarrow\infty}
\KLdiv{p^j_{xv}}{p_{xy}}=d\opt.
$$
In view of Lemma \ref{ottimofinito}, we can assume that all $p^j_{xvz}$ have finite mean vector $\mu_j$ and covariance matrix $\bar{K}_j$.
Let $m_j$ and $K_j$ be the mean and covariance of $\left[\begin{smallmatrix}x\\v\end{smallmatrix}\right]$, i.e.  $m_j$ are the first $2n$ components of $\mu_j$ and $K_j$ is the $2n\times 2n$ upper-left block of $\bar{K}_j$.
Now notice that, as $j\rightarrow \infty$, $\|K_j\|$ and $\|m_j\|$ remain bounded.
In fact, in view of Lemma \ref{lem:opt_gauss},
\begin{equation}\label{eqdivergente}\begin{split}
\KLdiv{p^j_{xv}}{p_{xy}}\geq \KLdiv{p^{Gj}_{xv}}{p_{xy}} = \\
\trace[K_{\bigl[\begin{smallmatrix}x\\ y\end{smallmatrix}\bigr]}^{-1}K_j] +m_j^\ast K_{\bigl[\begin{smallmatrix}x\\ y\end{smallmatrix}\bigr]}^{-1} m_j-\ln\left[
\frac{\det[K_j]}{\det[K_{\bigl[\begin{smallmatrix}x\\ y\end{smallmatrix}\bigr]}]}\right]-2n,
\end{split}\end{equation}
where $p^{Gj}_{xv}$ is the Gaussian distribution having mean vector $m_j$ and covariance matrix $K_j$.
It is easy to check that the right-hand side of (\ref{eqdivergente}) diverges if at least one of $\|K_j\|$ and $\|m_j\|$ does. Hence, both $\|K_j\|$ and $\|m_j\|$ remain bounded. Thus, also $\mu_j$ and  $\bar{K}_j$
remain bounded.
Therefore, there exists a subsequence $p^{j_i}_{xvz}$ such that $\bar{K}_{j_i}$ and $\mu_{j_i}$ converge.
Let $\bar{K}\opt$ and $\mu\opt$ be their limits and let $K\opt$ and $m\opt$ be the limits of 
$K_{j_i}$ and $m_{j_i}$.
Notice now that each density of the corresponding sequence $p^{Gj_i}_{xvz}$ satisfies the constraints of Problem \ref{prob:preci-math-prob}. In fact, the marginal $p_{xz}$ does not change and, in view of \eqref{eq:K_xvz_inv}, the second constraint only depends on the variance matrix.
Therefore, also the Gaussian distribution $p^{G\star}_{xvz}$, whose mean and variance are $\bar{K}\opt$ and $\mu\opt$, satisfies the constraints of Problem \ref{prob:preci-math-prob}. Let $p^{G\star}_{xv}$ be the corresponding marginal. We have
\bea\nonumber
d\opt&=&\lim_{i\rightarrow\infty}
\KLdiv{p^{j_i}_{xv}}{p_{xy}}\geq \lim_{i\rightarrow\infty} \KLdiv{p^{Gj_i}_{xv}}{p_{xy}}
\\\nonumber
&=&
\lim_{i\rightarrow\infty}\trace[K_{\bigl[\begin{smallmatrix}x\\ y\end{smallmatrix}\bigr]}^{-1}K_{j_i}] +m_{j_i}^\ast K_{\bigl[\begin{smallmatrix}x\\ y\end{smallmatrix}\bigr]}^{-1} m_{j_i}- \\
& & + \ln\left[
\frac{\det[K_{j_i}]}{\det[K_{\bigl[\begin{smallmatrix}x\\ y\end{smallmatrix}\bigr]}]}\right]-2n\\
\nonumber&=&
\trace[K_{\bigl[\begin{smallmatrix}x\\ y\end{smallmatrix}\bigr]}^{-1}K\opt] +(m\opt)^\ast K_{\bigl[\begin{smallmatrix}x\\ y\end{smallmatrix}\bigr]}^{-1} m\opt- \\
& & + \ln\left[
\frac{\det[K\opt]}{\det[K_{\bigl[\begin{smallmatrix}x\\ y\end{smallmatrix}\bigr]}]}\right]-2n\\
&=& 
\KLdiv{p^{G\star}_{xv}}{p_{xy}}.
\label{eq:finlpro}
\eea
Thus $p^{G\star}_{xvz}$ solves Problem \ref{prob:preci-math-prob}.

\end{IEEEproof}

Notice that from (\ref{eq:finlpro}) it is immediate to see that the optimal solution not only exists but is Gaussian distributed with zero mean.
}
{
\begin{corollary}
Let $x$ and $y$ be jointly Gaussian. Then the solution of Problem \ref{prob:preci-math-prob} is  zero mean and Gaussian. 
\end{corollary}}
{We are now ready to find the solution of our problem}.
\begin{theorem}\label{prop:p_xv_opt}
{
The solution of Problem \ref{prob:preci-math-prob} is the zero mean circular symmetric Gaussian density $p\opt_{xvz}$} whose covariance matrix is
\begin{equation}
\label{soluzottimaformacov}
	K_{\bigl[\begin{smallmatrix}x\\v\\z\end{smallmatrix}\bigr]}(Z,C) = \begin{bmatrix} K_{xx} & K_{xz}K_{zz}^{-1}Z^*&K_{xz}\\ ZK_{zz}^{-1}K_{xz}^* & Z K_{zz}^{-1} Z^* + CC^*&Z\\
	K_{xz}^\ast& Z^\ast &K_{zz}
	\end{bmatrix},
\end{equation}
where $Z$ and $C$ solve 
\begin{equation}
\label{eq:CZ_opt}
	\begin{cases}
        C^*= C^*(Z K_{zz}^{-1}B K_{zz}^{-1}Z^* + CC^*)^{-1}A \\
         Z^* = K_{zx}K_{xx}^{-1}K_{xy}+ BK_{zz}^{-1}Z^*(ZK_{zz}^{-1}BK_{zz}^{-1}Z^*+ \\
         \quad + CC^*)^{-1}A
        \end{cases}
\end{equation}	
with
\begin{gather}
	A:= K_{yy} - K_{xy}^*K_{xx}^{-1}K_{xy},\\
	B:= K_{zz} - K_{xz}^*K_{xx}^{-1}K_{xz}.
\end{gather}
\end{theorem}
\begin{IEEEproof}
See the Appendix.
\end{IEEEproof}

{In view of \eqref{KLq2} and \eqref{boundint}, Theorem \ref{prop:p_xv_opt} provides the tightest bound to the error region \eqref{boundint}. 
Indeed, let $K_{\bigl[\begin{smallmatrix}x\\v\end{smallmatrix}\bigr]}$ be a shorthand notation for the $2n\times 2n$ upper-left corner of (\ref{soluzottimaformacov}).
Then, $D\opt$ is given by 
\begin{equation}
\begin{split}
	D\opt = \mathbb{D}(p\opt_{xv}\|p_{xy})= -\log\det(K_{\bigl[\begin{smallmatrix}x\\v\end{smallmatrix}\bigr]}	K_{\bigl[\begin{smallmatrix}x\\y\end{smallmatrix}\bigr]}^{-1}) + \\
	+ \trace{K_{\bigl[\begin{smallmatrix}x\\y\end{smallmatrix}\bigr]}}^{-1}\left(K_{\bigl[\begin{smallmatrix}x\\v\end{smallmatrix}\bigr]}-K_{\bigl[\begin{smallmatrix}x\\y\end{smallmatrix}\bigr]}\right).
	\end{split}
\end{equation}

Consider the circular symmetric Gaussian density $p\opt_{xvz}$, with zero mean and variance
\begin{equation}
	K_{\Bigl[\begin{smallmatrix}x\\v\\z\end{smallmatrix}\Bigr]} =
	\begin{bmatrix}
		K_{xx}	&	K_{xz}K_{zz}^{-1}Z^*	& K_{xz}\\ 		   ZK_{zz}^{-1}K_{xz}^*	& Z K_{zz}^{-1} Z^* + CC^*	& Z\\
		K_{xz}^*	&	Z^*	&	K_{zz}
	\end{bmatrix}.
\end{equation}
Note that it is such that $x$ and $v$ are conditionally independent given $z$. Then, by marginalizing and conditioning, we can obtain an optimum {attacking} strategy $p\opt_{v|z}(\cdot|a)$ which achieves \eqref{boundint}. It is given by the proper Gaussian density whose mean and variance are defined by
\begin{align}
	 \mu_{v|z} & := ZK_{zz}^{-1}a\\
	 K_{v|z} & := K_{vv} - K_{vz}K_{zz}^{-1}K_{vz}^* = CC^*
\end{align}

  
{
}

\section{Efficient computation of the tightest bound}\label{sec:algorithm}
In view of Theorem \ref{prop:p_xv_opt}, in order to provide the expression of the optimal solution $p\opt_{xvz}$, we have to compute matrices $C,Z$ which solve the system of nonlinear matrix equations \eqref{eq:CZ_opt}.  This appears however to be a highly non trivial task. Thus, we propose a two stage algorithm:
\begin{enumerate}
\item {\bf Feasible (projected) Solution.} To begin with, we deal with an optimization problem which can be considered a relaxed version of Problem \ref{prob:preci-math-prob}, since no positivity constraints on the matrix $K_{\Bigl[\begin{smallmatrix}x\\v\\z\end{smallmatrix}\Bigr]}$ are imposed. This task turns out to be much simpler to achieve. Indeed, the solution can be computed in closed form.  Then, we project the solution to the relaxed problem onto the feasible set, i.e. the set of pairs $(X,Z)$ which make $K_{\Bigl[\begin{smallmatrix}x\\v\\z\end{smallmatrix}\Bigr]}$ positive definite.
\item {\bf Iterative Algorithm.} We use the projection as a starting point for an iterative update procedure whose fixed point satisfies \eqref{eq:CZ_opt}.
\end{enumerate}

Next we provide some details for each phase.

\noindent {\bf Feasible Solution.}  Minimizing \eqref{KLq2} with no constraints on the positivity of $K_{\Bigl[\begin{smallmatrix}x\\v\\z\end{smallmatrix}\Bigr]}$ is equivalent to solve
\begin{problem}\label{prob:opt_no_pos}
\begin{equation}
\begin{split}
\arg\min_{X,Z} J(
K_{\bigl[\begin{smallmatrix}x\\v\end{smallmatrix}\bigr]}(Z,X)
):= &  \left\lbrace{-\log\det(
K_{
\bigl[
\begin{smallmatrix}x\\v\end{smallmatrix}
\bigr]
}(Z,X)
K_{
\bigl[
\begin{smallmatrix}x\\y\end{smallmatrix}
\bigr]}
^{-1}) + }\right.\\
& \left. \trace K_{
\bigl[
\begin{smallmatrix}x\\y\end{smallmatrix}
\bigr]
}^{-1}
K_{\bigl[
\begin{smallmatrix}x\\v\end{smallmatrix}
\bigr]
}
\right\rbrace
\end{split}
\end{equation}
\end{problem}
where
\begin{gather}
K_{\bigl[\begin{smallmatrix}x\\v\end{smallmatrix}\bigr]}(Z,X):= 
\begin{bmatrix} K_{xx} &K_{xz}K_{zz}^{-1}Z^*\\ Z(K_{xz}K_{zz}^{-1})^* &X \end{bmatrix}, \\
K_{\bigl[\begin{smallmatrix}x\\y\end{smallmatrix}\bigr]}:=
\begin{bmatrix} K_{xx} & K_{xy}\\ K_{xy}^* & K_{yy} \end{bmatrix}. 
\end{gather}
In the same vein of the proof of Theorem \ref{prop:p_xv_opt}, we work out the optimality conditions that $X$ and $Z$ have to satisfy, based on the analysis of the first variation $D[J(K_{\bigl[\begin{smallmatrix}x\\v\end{smallmatrix}\bigr]}(Z,X);\delta K_{\bigl[\begin{smallmatrix}x\\v\end{smallmatrix}\bigr]}]$. Some easy algebraic calculations lead us to the closed form of an optimal solution $(Z,X)$:
\begin{equation}\label{eq:optXZ_no_pos}
\begin{cases}
Z = K_{xy}^*{K_{xx}}^{-1}K_{xz}(K_{xz}^*K_{xx}^{-1}K_{xz})^\dagger K_{zz},\\
X = K_{yy} - K_{xy}^* {K_{xx}}^{-\frac{1}{2}} \left[I - \right. \\ \left. \quad K_{xx}^{-\frac{1}{2}}K_{xz}(K_{xz}^*{K_{xx}}^{-1}K_{xz})^\dagger K_{xz}^*K_{xx}^{-\frac{1}{2}}\right]{K_{xx}}^{-\frac{1}{2}}K_{xy}
\end{cases}
\end{equation}
where ``~$^\dag$~'' denotes Moore-Penrose pseudo inverse.

{If the obtained $X$ and $Z$ are such that $X-ZK_{zz}^{-1}K^*\geq 0$, the algorithm terminates. Otherwise, a pair $(C,Z)$ is obtained as follows. Let $T$ be a unitary matrix such that $\Sigma_T:=T^\ast(X-ZK_{zz}^{-1}K^*)T={\rm diag}(d_1,d_2,\dots,d_k,\delta_1,\delta_2,\dots,\delta_h)$, where $d_i$ are positive and in decreasing order, and $\delta_i$ are negative or zero. 
Let $\Sigma_T':={\rm diag}(d_1,d_2,\dots,d_k,\varepsilon,\varepsilon,\dots,\varepsilon)$, where $\varepsilon:=(d_k/100)>0$ is a ``small" parameter.
Let $\Sigma':=T\Sigma_T'T^\ast >0$ and $C$ be such that $CC^*=\Sigma'$.

\noindent {\bf Iterative Algorithm.} We use the pair $(C,Z)$ as a}  starting point for the iterations
\begin{equation}
\label{eq:iterations}
	\begin{cases}
        C^*(k+1) = C^*(k)(Z(k) K_{zz}^{-1}B K_{zz}^{-1}Z^*(k) + C(k)C^*(k))^{-1}A, \\
         Z^*(k+1) = K_{zx}K_{xx}^{-1}K_{xy}+ \\ \quad BK_{zz}^{-1}Z^*(k)(Z(k)K_{zz}^{-1}BK_{zz}^{-1}Z^*(k)+C(k)C^*(z))^{-1}A
        \end{cases}
\end{equation}	
where 
\begin{gather}
	A:= K_{yy} - K_{xy}^*K_{xx}^{-1}K_{xy}\\
	B:= K_{zz} - K_{xz}^*K_{xx}^{-1}K_{xz}.
\end{gather}
By the the iterative process we aim at finding a fixed point for \eqref{eq:iterations}, which provides the solution of Problem \ref{prob:preci-math-prob}. The iterative process can be stopped either after a fixed number of iterations, or when the variation of $D^*$ over one iteration is smaller than a given percentage.

\section{Numerical results}\label{sec:numerical_results}

\subsection{Uncorrelated Channels}

In order to assess the performance of the proposed algorithm for the computation of the tightest bound, we first consider the case where 
$m = n$ and the covariance matrices are identities, i.e. $$ K_{\bigl[\begin{smallmatrix}x\\y\\z\end{smallmatrix}\bigr]} =
\left[\begin{array}{ccc}
I_n & \sigma I_n & \rho I_n \\
\sigma^\ast I_n &  I_n & \tau I_n \\
 \rho^\ast I_n & \tau^\ast I_n &  I_n
\end{array} \right]$$
This scenario corresponds for example to an \ac{OFDM} transmission with uncorrelated channel frequency response. Beyond being an asymptotic case widely considered in the literature, this is also a practical scenario, when a subset of subcarriers with cardinality smaller than the number of channel taps is considered, and the channel taps are independent Gaussian variables. 
The parameter $\rho$ dictates the correlation between  channel estimates performed by Eve and the legitimate channel.

\begin{figure}
\centering
\includegraphics[width=1\hsize]{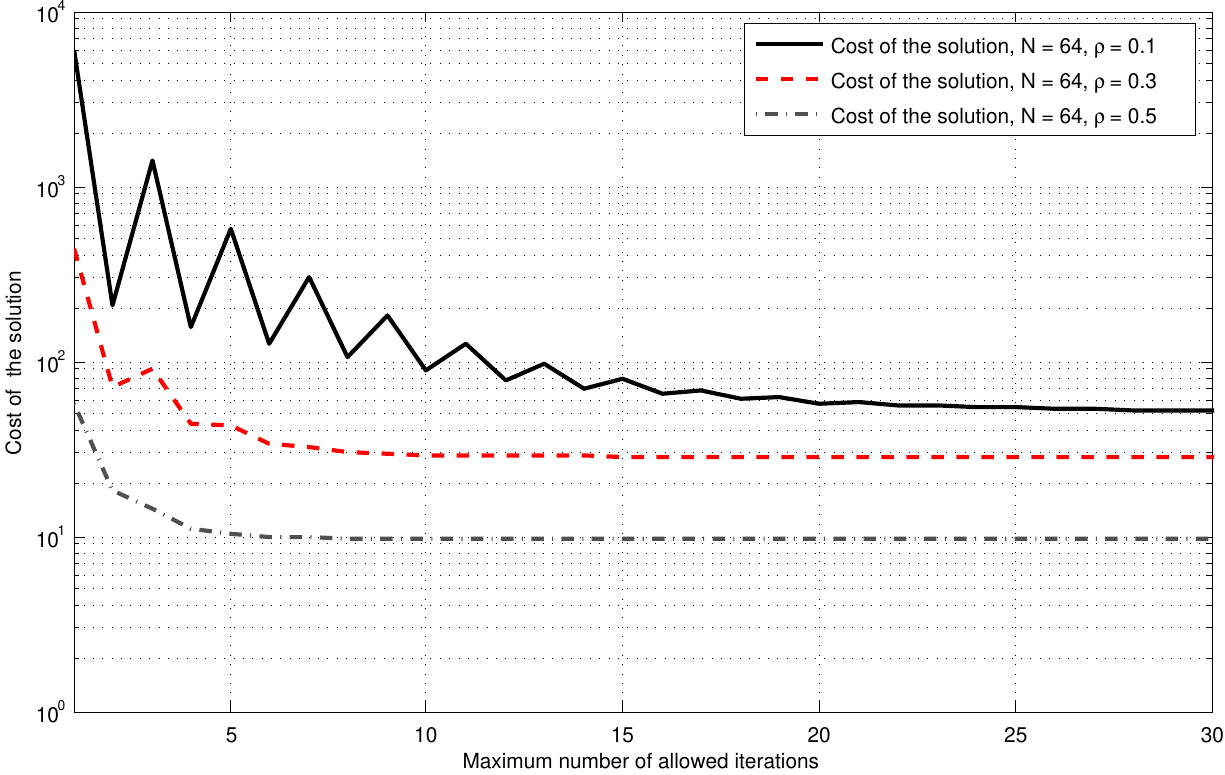}
\caption{Cost of the solution computed by the iterative algorithm as a function of the maximum number of iterations, with $n=m=64$, for $\rho = 0.1, 0.3, 0.5$.}
\label{fig0}
\end{figure}
First we assess the performance of the iterative algorithm. Fig \ref{fig0} shows the values of the cost of the optimum solution $D^*$ as a function of the number of iterations for the iterative algorithm, with $n=m=64$, and various values of $\rho$.  {We observe that the iterative algorithm always converges to a fixed point for \eqref{eq:iterations} and that the convergence to a solution with good accuracy takes less than $100$ iterations. Thus, in the following we consider this value for the maximum number of iterations}.
\begin{figure}
\centering
\includegraphics[width=1\hsize]{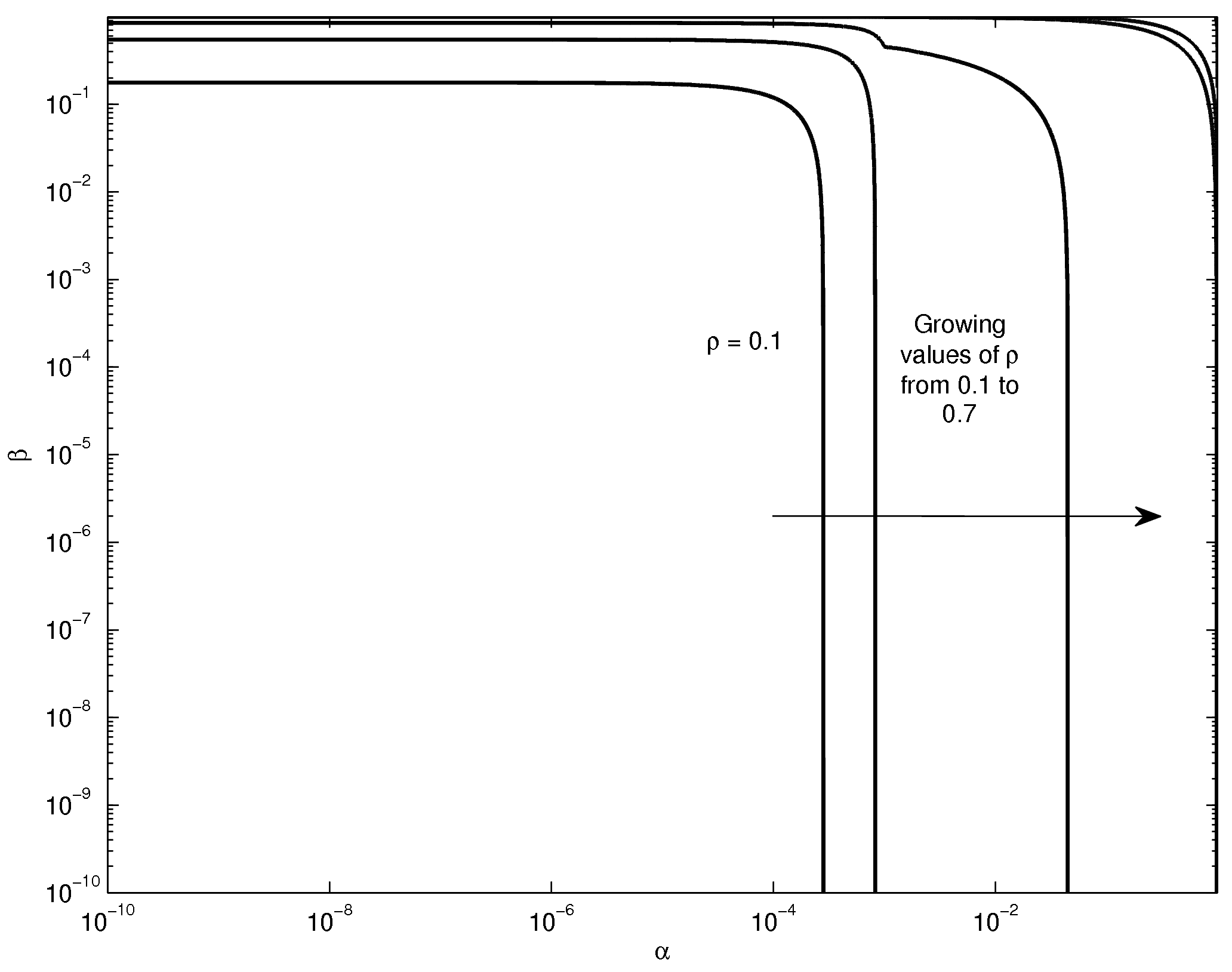}
\caption{Bound of the region type II ($\beta$) vs type I ($\alpha$) error probability for various values of the correlation parameter $\rho$, with  $K_{xx} = I_{n \times n}$, $K_{zz} = I_{m \times m}$, and $K_{xz} = \rho I_{n \times m}$.}
\label{fig1}
\end{figure}

Fig. \ref{fig1} shows the bound of the type II ($\beta$) -- type I ($\alpha$) error probability region for various values of the correlation parameter $\rho$, and for $n=m=64$, as obtained from the proposed iterative approach. As expected, we observe that for increasing values of $\rho$, the region of achievable values of $\alpha$ and $\beta$ gets wider. In particular, for the considered scenario, the type II error probability is larger than $10^{-1}$ already for $\rho=0.4$.
\begin{figure}
\centering
\includegraphics[width=1\hsize]{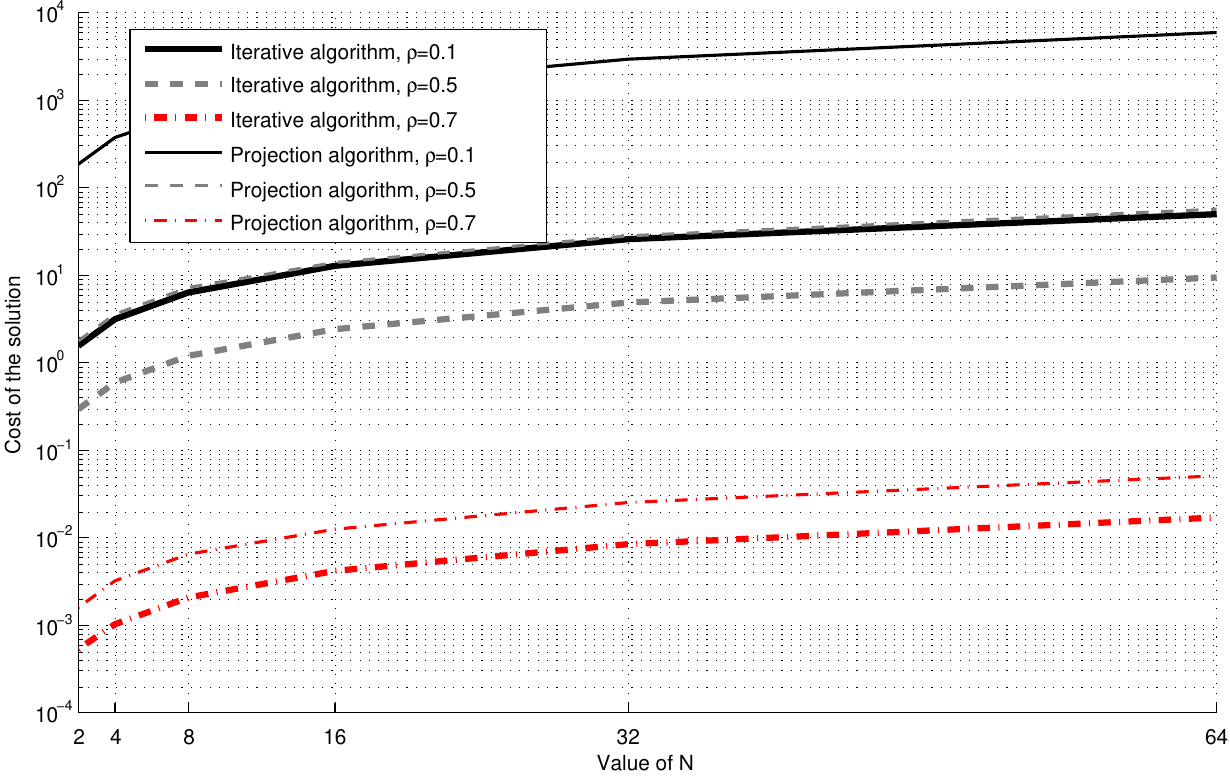}
\caption{Cost function $D^*$ as a function of $n$ for various values of the correlation parameter $\rho$, with  $K_{xx} = I_{n \times n}$, $K_{zz} = I_{m \times m}$, and $K_{xz} = \rho I_{n \times m}$. Both projection and iterative algorithms are considered.}
\label{fig2}
\end{figure}

In Fig. \ref{fig2} we report the results obtained for both the initial feasible solution (projection of the solution of \eqref{eq:optXZ_no_pos}) and final solution of the iterative algorithm, as a function of $n$, for~$\rho=0.1$,$0.5$,$0.7$. {For the sake of clarity, we also show the cost of the solutions provided by the iterative algorithm in Tab. \ref{tab:cost_iterative_solutions}}. 

\begin{table}
\caption{Cost of the solution provided by 
the iterative solution.}
\centering
\begin{tabular}{c|cccccc}
\hline
$D^*$ & $n = 2$ & $n = 4$ & $n = 8$ & $n = 16$ & $n = 32$ & $n = 64$ \\ 
\hline
$\rho$ = $0.1$ & $1.6099$ & $3.2199$ & $6.4397$ & $12.8795$ & $25.7589$ & 	$51.5179$ \\  
$\rho$ = $0.5$ & $0.3047$ & $0.6094$ & $1.2189$ & $2.4378$ & $4.8756$ & $9.7511$ \\ 
$\rho$ = $0.7$ & $0.0005$& $0.0011$ & $0.0021$ & $0.0042$ & $0.0085$ & $0.0169$ \\ 
\hline
\end{tabular} 
\label{tab:cost_iterative_solutions}
\end{table}
 We note that the iterative algorithm remarkably lowers the value of the cost function from the initial feasible solution, thus motivating its use, although it comes at a cost of more computations. Also, as expected, the cost function increases with $n$. For the considered case of \ac{OFDM} transmission, this means that more dispersive channels having independent taps provide potentially a better authentication system. This phenomenon has been already seen in \cite{Baracca12}.

\subsection{Correlated Channels}

\begin{figure}
\subfloat[][\emph{Iterative algorithm}.]
{\includegraphics[width=1\hsize]{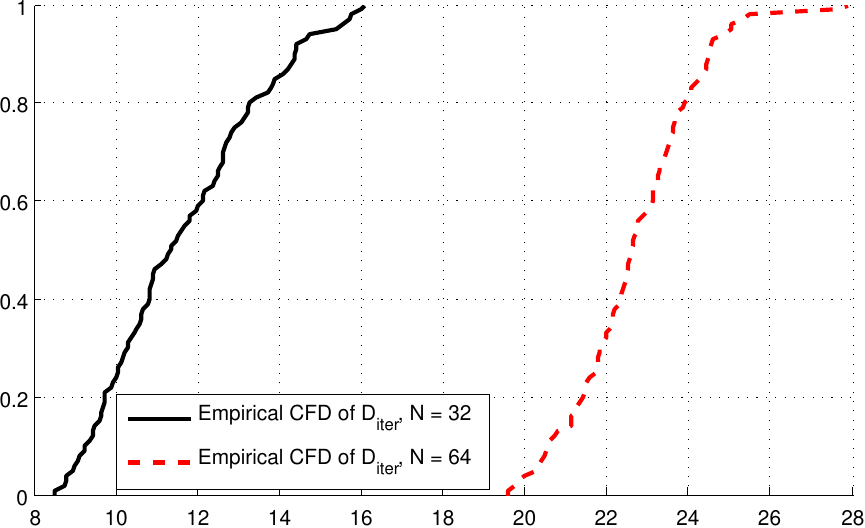}} \quad
\subfloat[][\emph{Initial feasible solution}.]
{\includegraphics[width=1\hsize]{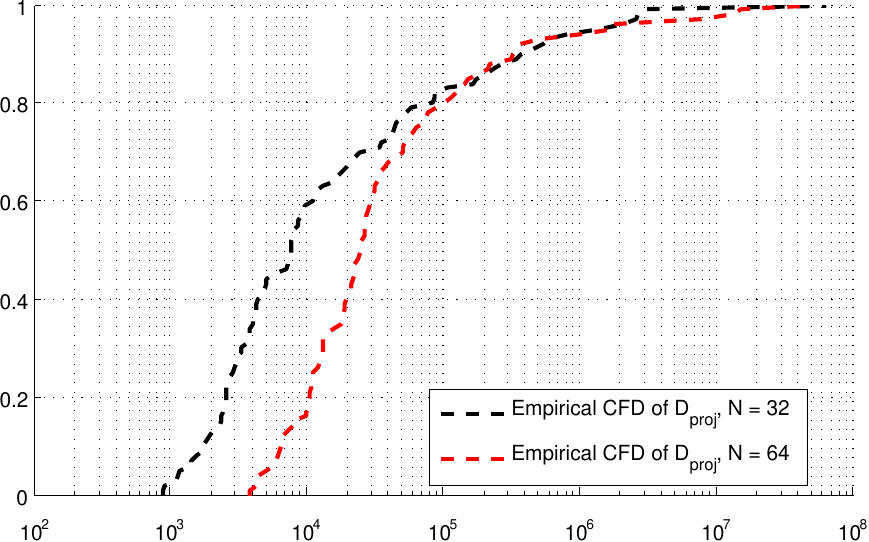}} \\
\caption{CDF of the cost function for two values of $n$.}
\label{fig3}
\end{figure}

We now consider channels with random correlation.  
We let $m = n$ and generate $K_{\Bigl[\begin{smallmatrix}x\\y\\z\end{smallmatrix}\Bigr]}$ as a realization of a $3n\times 3n$ real Wishart matrix\footnote{A $n\times n$ real (resp., complex) \emph{Wishart matrix} is a random matrix $W$ that can be written as $W = AA^\ast$, where $A$ is a $n\times n$ random matrix with \ac{iid} real (resp., circularly symmetric complex) Gaussian entries. In our case, the entries of $A$ have zero mean and unit variance.}. \nota{Perch\'e reale? Non dovremmo usare matrici complesse? Ovviamente non mi aspetto cambiamenti nei risultati, ma \`e solo per coerenza con il `system model'}}
Even in this case we verified that setting the maximum number of iteration to 100 is enough for the convergence of the iterative algorithm. Fig.~\ref{fig3}.a shows the cumulative distribution function (CDF) of $D^*$ for two values of $n=m$, at the convergence of the iterative algorithm. Also in this case we observe that a larger $n$ provides a larger value of $D^*$. We also report in Fig. \ref{fig3}.b the CDF for the initial feasible solution obtained by projection.


For the random correlation case, 
Tab. \ref{tab:positive}
shows the probability that the closed form solution of the relaxed problem \eqref{eq:optXZ_no_pos} satisfies the positivity constraint, as a function of $n$. 

\begin{table}
\caption{Probability that \eqref{eq:optXZ_no_pos} is feasible, as a function of $n$.}
\centering
\begin{tabular}{c|cccccc}
\hline
 $n$ &  $2$ & $4$ & $8$ & $16$ & $32$ & $64$ \\ 
\hline
$p$ $[\%]$ & $43$ & $10$ & $0$ & $0$ & $0$ & $0$  \\
\hline
\end{tabular}
\label{tab:positive} 
\end{table}

Note that as $n$ increases this probability goes fast to zero, thus making the projection step necessary to obtain an initial feasible solution for the iterative algorithm.

\begin{figure}
\centering
\includegraphics[width=1\hsize]{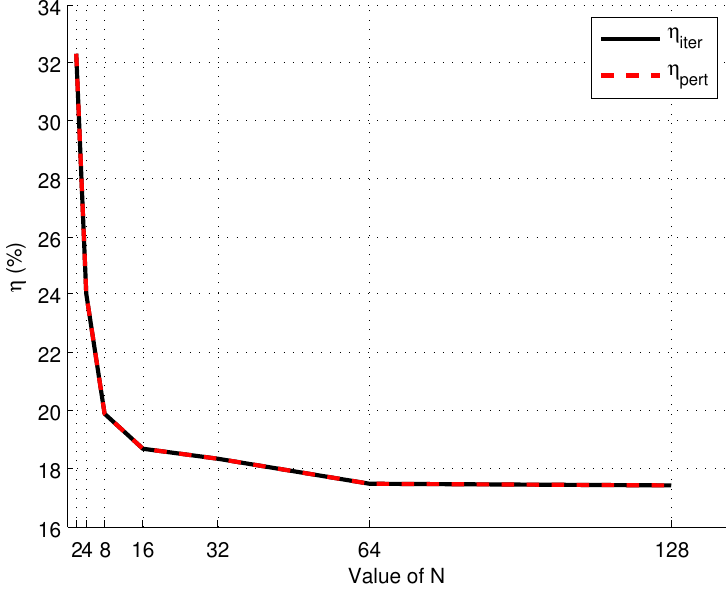}
\caption{Percentage improvement $\eta$ as a function of $n$. Random correlation matrices and $n=m$. Perturbation analysis results are included.}
\label{fig5}
\end{figure}

{In order to compare the iterative solution to the one provided by \eqref{eq:optXZ_no_pos}, which may not fulfill the positivity constraints on the joint covariance matrix, Fig. \ref{fig5} shows {the percentage increase of the cost \eqref{eq:J} defined as
\begin{equation}
\eta := 100 \times \left[\frac{J^*_{iter}}{J^*_{cf}} - 1\right]\,,
\end{equation}
where $J^*_{iter}$ is the cost of the solution provided by the iterative algorithm, whereas $J^*_{cf}$  is the cost of the one computed in closed form through \eqref{eq:optXZ_no_pos}.}  
The analysing of the increment with regard to $J^*$ is convenient because $D^*_{cf}$ can vanish. Indeed, recall that, if $K_{\bigl[\begin{smallmatrix}x\\y\end{smallmatrix}\bigr]}$  is a $n\times n$ matrix, it holds that $D^*= J^*-2n$. 
We note that the increase is in the range of 20\% to 30\% for the considered scenario}. {Moreover, it is diminishing as $n$ increases. This seems to suggest that, for growing values of $n$, the solution computed by means of \eqref{eq:optXZ_no_pos} corresponds to a matrix of the form \eqref{eq:matrixK(XZ)} which gets closer to the cone of positive definite matrices of size $(2n+m)$.}

We also provide results for the perturbation analysis. In particular, we evaluate the effects of small perturbations of $Z$ and $C$ generated as Gaussian random variables with norm $0.01\|Z\|$ and $0.01\|C\|$, respectively. Fig. \ref{fig5} reports the maximum cost function achieved for all perturbed values, showing that it provides negligible improvement with respect to the solution of the iterative approach. This supports the conclusion that the iterative approach reaches a minimum point for $\mathbb{J}(K_{\bigl[\begin{smallmatrix}x\\v\end{smallmatrix}\bigr]}(Z,C))$. We also applied the iterative algorithm starting from the perturbed solutions which led to cost improvements. Results, not reported here, show that this procedure achieves very small improvements with an increase of the cost function of $0.01$\% . 

\section{Conclusions}\label{sec:conc}

We have considered the problem of deriving a universal performance bound, for a message source authentication scheme based on channel estimates in a wireless fading scenario, where an attacker may have correlated observations available.
We have formulated an outer bound to the region of achievable false alarm and missed detection probabilities, which is universal across all possible decision rules by the receiver.

Under the assumption that the channels are represented by multivariate complex Gaussian variables, we have proved that the tightest bound corresponds to a forging strategy that produces a zero mean signal which is jointly Gaussian with the attacker observations. Furthermore, we have derived a characterization of their joint covariance matrix {through the solution of a system of two nonlinear matrix equations}.
Based upon this characterization, we have also devised an efficient iterative algorithm for its computation: the solution to the matrix system appears as fixed point of the iteration.

From numerical results, we conjecture that the proposed iterative approach for the best {attacking} strategy converges in general, although determining its convergence seems a highly difficult problem. Moreover, from the perturbation analysis, we deduce that the limit point is a local minimum. We have therefore provided an effective method for the {attacking} strategy that yields the tightest bound on the error region of any message authentication procedure.

\appendix

In this Appendix we provide the proof of Theorem 2.

We have already shown that the optimal solution is a zero-mean Gaussian distribution having covariance matrix of the form
\begin{equation}
K_{\Bigl[\begin{smallmatrix}x\\v\\z\end{smallmatrix}\Bigr]}=\begin{bmatrix} K_{xx} & Y & K_{xz}\\ Y^* & X & Z \\ K_{xz}^* & Z^* & K_{zz}  \end{bmatrix},
\end{equation}
where 
$$
K_{\bigl[\begin{smallmatrix}x\\z\end{smallmatrix}\bigr]}:=\begin{bmatrix} K_{xx} & K_{xz}\\ K_{xz}^* & K_{zz}  \end{bmatrix} > 0
$$
is given.
Clearly in this way the first constraint of Problem \ref{prob:preci-math-prob} is automatically satisfied for any $X,Y,Z$.
We now show that the second constraint is equivalent to impose
$$Y=K_{xz}K_{zz}^{-1}Z^\ast.$$
Indeed, 
in view of Lemma \ref{lem:cond_ort}, $x$ and $v$ are conditional orthogonal given $z$, so that the inverse of $K_{xvz}$ must exhibit the zero-block pattern \eqref{eq:K_xvz_inv}.
Based on this information, we can compute $Y$ as a function of $Z$ and $X$ by employing the block-matrix inversion formula in (\ref{eq:block_inversion}) at the top of the page
\begin{figure*}[t]
\begin{equation}\label{eq:block_inversion}
    M_1 = \begin{bmatrix} A_1 & B_1 \\ C_1 & D_1 \end{bmatrix} \, 
        \Rightarrow \,
    M_1^{-1} = \begin{bmatrix} {(A_1-B_1D_1^{-1}C_1)}^{-1} & -A_1^{-1}B_1{(D_1-C_1A_1^{-1}B_1)}^{-1} \\ - D_1^{-1}C_1{(A_1-B_1D_1^{-1}C_1)}^{-1} & {(D_1-C_1A_1^{-1}B_1)}^{-1} \end{bmatrix}.
\end{equation} 
\hrulefill
\vspace*{4pt}
\end{figure*}
We partition $K_{\Bigl[\begin{smallmatrix}x\\v\\z\end{smallmatrix}\Bigr]}$
as 
\be\label{parcovsoltbt}
K_{\Bigl[\begin{smallmatrix}x\\v\\z\end{smallmatrix}\Bigr]}= \begin{bmatrix} A_1 & B_1 \\ C_1 & D_1 \end{bmatrix}, 
\ee
where
\begin{gather*}
A_1:= K_{xx},\quad 
B_1:= \begin{bmatrix} Y & K_{xz}\end{bmatrix},\\
C_1:= \begin{bmatrix} Y^* \\ K_{xz}^*\end{bmatrix},  \quad
D_1:= \begin{bmatrix} X & Z \\ Z^* &K_{zz}\end{bmatrix}.
\end{gather*}
Therefore, the block in position $(1,2)$ of $K_{xvz}^{-1}$  (with respect to the partition (\ref{parcovsoltbt})) is given by 
\begin{equation*}
    \begin{split}
     &-A_1^{-1}B_1{(D_1-C_1A_1^{-1}B_1)}^{-1} = -{K_{xx}}^{-1}\begin{bmatrix} Y  K_{xz}\end{bmatrix} 
                                     \times \\ 
                  &                   {\left({\begin{bmatrix} X & Z \\ Z^* & K_{zz} \end{bmatrix} - \begin{bmatrix} Y^* \\ K_{zx}\end{bmatrix}{K_{xx}}^{-1}\begin{bmatrix} Y & K_{xz}\end{bmatrix}}\right)}^{-1}\\
                                    &= -{K_{xx}}^{-1}\begin{bmatrix} Y & K_{xz}\end{bmatrix} \times \\
                                    &
                                    {\left(\underbrace{{\begin{bmatrix} X - Y^*{K_{xx}}^{-1}Y & Z - Y^*{K_{xx}}^{-1}K_{xz} \\ 
                                                            Z^*-K_{zx}K_{xx}^{-1}Y & K_{zz}-K_{zx}K_{xx}^{-1}K_{xz} \end{bmatrix} }}_{:= M_2}\right)} ^{-1}.
    \end{split}
\end{equation*}
In order to impose the zero-block pattern \eqref{eq:K_xvz_inv} to the inverse, we make the block in position $(1,1)$ in $-A_1^{-1}B_1{(D_1-C_1A_1^{-1}B_1)}^{-1}$ vanish. Note that we need to compute explicitly only the elements in the first column block of ${M_2}^{-1}$.
Let
\begin{equation}
\begin{split}
\begin{bmatrix} 
            A_2 & B_2\\ 
            C_2 & D_2 
      \end{bmatrix}  := \begin{bmatrix} 
            X - Y^*{K_{xx}}^{-1}Y & Z - Y^*{K_{xx}}^{-1}K_{xz} \\ 
            Z^*-K_{zx}K_{xx}^{-1}Y & K_{zz}-K_{zx}K_{xx}^{-1}K_{xz} 
      \end{bmatrix} \\
      = M_2 
\end{split}
\end{equation}
Thus, in view of the matrix inversion lemma, the first column block in $M_2^{-1}$ is given by
$$
\begin{bmatrix} 
            {(A_2 - B_2{D_2}^{-1}C_2)}^{-1}\\ 
            -{D_2}^{-1}C_2{(A_2-B_2D_2^{-1}C_2)}^{-1} 
      \end{bmatrix} .
$$
Therefore, orthogonality of $x$ and $v$ given $z$ implies
\begin{equation*}
    \begin{split}
        0 &= -{K_{xx}}^{-1}\begin{bmatrix} Y & K_{xz}\end{bmatrix}\begin{bmatrix} 
            {(A_2 - B_2{D_2}^{-1}C_2)}^{-1}\\ 
            -{D_2}^{-1}C_2{(A_2-B_2D_2^{-1}C_2)}^{-1} 
      \end{bmatrix} \\
          &= -{K_{xx}}^{-1}Y{(A_2 - B_2{D_2}^{-1}C_2)}^{-1} + \\
          & {K_{xx}}^{-1}K_{xz}{D_2}^{-1}C_2{(A_2-B_2D_2^{-1}C_2)}^{-1}\\
          &=  Y - K_{xz}{D_{2}}^{-1}C_2,
    \end{split}
\end{equation*}
so that 
\begin{equation*}
\begin{split}
    Y =  & K_{xz}\left({K_{zz}-K_{zx}K_{xx}^{-1}K_{xz}}\right)^{-1} \left(Z^*-K_{zx}K_{xx}^{-1}Y\right)\\
      = &\left[ \left( I+K_{xz}\left({K_{zz}-K_{zx}K_{xx}^{-1}K_{xz}}\right)^{-1} \right.\right.\times \\
      & \left.\left.K_{zx}{K_{xx}}^{-1}\right) \right]^{-1}K_{xz}\left(K_{zz}-K_{zx}K_{xx}^{-1}K_{xz}\right)^{-1} Z^*\\
      =& K_{xz}K_{zz}^{-1}Z^*.
    \end{split}
\end{equation*}
In this way, we have parametrized all the matrices $K_{\Bigl[\begin{smallmatrix}x\\v\\z\end{smallmatrix}\Bigr]}$ whose inverse has the specified structure.
At this point, we could minimize the divergence $\mathbb{D}(p_{xv}\|p_{xy})$ over $Z$ and $X$. This turns out to be an easy problem that can be solved in closed form. This solution, however, is not the solution\footnote{Here we mention this simplified optimization problem because, as discussed later, it turns out to be very useful as the first step of an efficient numerical procedure that computes the solution of our original problem.} of our original problem since 
there is yet another (hidden) constraint that we need to impose. Namely we have to impose that the matrix 
\begin{equation}\label{eq:matrixK(XZ)}
K_{\Bigl[\begin{smallmatrix}x\\v\\z\end{smallmatrix}\Bigr]}=\begin{bmatrix} K_{xx} & K_{xz}K_{zz}^{-1}Z^\ast & K_{xz}\\ (K_{xz}K_{zz}^{-1}Z^\ast)^* & X & Z \\ K_{xz}^* & Z^* & K_{zz}  \end{bmatrix}
\end{equation}
is a {\em bona fide} covariance matrix, i.e. it is positive semidefinite. Since $K_{\bigl[\begin{smallmatrix}x\\z\end{smallmatrix}\bigr]}$ is positive definite, this constraint is equivalent to
$$
X-\begin{bmatrix} 
 (K_{xz}K_{zz}^{-1}Z^\ast)^\ast & Z\end{bmatrix}
\begin{bmatrix} K_{xx} & K_{xz}\\ K_{xz}^* & K_{zz}  \end{bmatrix}^{-1}
\begin{bmatrix}  K_{xz}K_{zz}^{-1}Z^\ast \\  Z^*  \end{bmatrix}
\geq 0
 $$
which, with simple algebraic manipulations, is seen to be equivalent to
\begin{equation}\label{eq:xz_pd}
X - Z K_{zz}^{-1} Z^*\geq 0.
\end{equation}
The positivity constraint is then automatically satisfied if we re-parametrize the unknown matrix $X$ in term of a new matrix $C$ in the form
\begin{equation}\label{eq:CC^*}
X = Z K_{zz}^{-1} Z^* + CC^*.
\end{equation} 
The optimal solution can be now easily obtained by solving the following {\em unconstrained} optimization problem
 \begin{equation}\label{eq:optimize_in_CZ}
\arg\min_{C,Z} \mathbb{D}(p_{xv}\|p_{xy}).
\end{equation}
Since
\begin{gather}\label{eq:H}
K_{\bigl[\begin{smallmatrix}x\\v\end{smallmatrix}\bigr]}(Z,C):= \begin{bmatrix} K_{xx} \\ &K_{xz}K_{zz}^{-1}Z^*\\ Z(K_{xz}K_{zz}^{-1})^* & Z K_{zz}^{-1} Z^* + CC^*\end{bmatrix}, 
\\ 
K_{\bigl[\begin{smallmatrix}x\\y\end{smallmatrix}\bigr]}:=\begin{bmatrix} K_{xx} & K_{xy}\\ K_{xy}^* & K_{yy} \end{bmatrix},
\end{gather}
solving \eqref{eq:optimize_in_CZ} is equivalent to compute  
\begin{equation}\label{eq:argmin_J}
\begin{split}
\arg\min_{Z,C}\, \left\lbrace{-\log\det(K_{\bigl[\begin{smallmatrix}x\\v\end{smallmatrix}\bigr]}(Z,C){K_{\bigl[\begin{smallmatrix}x\\y\end{smallmatrix}\bigr]}}^{-1}) + }\right.\\ 
\left.{\trace{{K_{\bigl[\begin{smallmatrix}x\\y\end{smallmatrix}\bigr]}}^{-1}K_{xv}K_{\bigl[\begin{smallmatrix}x\\v\end{smallmatrix}\bigr]}(Z,C)}}\right\rbrace.
\end{split}
\end{equation}
We are then led to the formulation of Problem \ref{prob:preci-math-prob}.
Let 
\begin{figure*}[t]
\setcounter{tempequationcounter}{\value{equation}}
\begin{equation}
\setcounter{tempequationcounter}{\value{equation}}
\begin{split}
\label{myeqtop}
&D [J(K_{xv}(Z,C));\delta K_{xv}(Z,C)] \\
&= \trace{\left[ (- K_{xv}^{-1} + {K_{xy}}^{-1})\delta K_{xv} \right]}\\
&= \trace{\left[ \underbrace{(- K_{xv}^{-1} + {K_{xy}}^{-1})}_{=:\Delta}\begin{bmatrix} 0 & K_{xz}K_{zz}^{-1}\delta Z^* \\ \delta Z (K_{xz}K_{zz}^{-1})^* &\delta Z K_{zz}^{-1}Z^*+ZK_{zz}^{-1}\delta Z^* + \delta C C^* + C\delta C^*\end{bmatrix}\right]}\\
&= \trace{\left[ \begin{bmatrix} \Delta_{11} & \Delta_{12}\\ \Delta_{21}&\Delta_{22}\end{bmatrix}\begin{bmatrix} 0 & K_{xz}K_{zz}^{-1}\delta Z^* \\ \delta Z (K_{xz}K_{zz}^{-1})^* &\delta Z K_{zz}^{-1}Z^*+ZK_{zz}^{-1}\delta Z^* + \delta C C^* + C\delta C^*\end{bmatrix}\right]}\\
&= \trace{\begin{bmatrix} \Delta_{12}\delta Z (K_{xz}K_{zz}^{-1})^* & * \\ * & \Delta_{21}K_{xz}K_{zz}^{-1}\delta Z^* + \Delta_{22}\left[{\delta Z K_{zz}^{-1}Z^*+ZK_{zz}^{-1}\delta Z^* + \delta C C^* + C\delta C^*}\right]\end{bmatrix}}.
\end{split}
\end{equation}
\setcounter{equation}{\value{tempequationcounter}}
\hrulefill
\vspace*{4pt}
\end{figure*}\begin{equation}\label{eq:J}
\begin{split}
J(K_{\bigl[\begin{smallmatrix}x\\v\end{smallmatrix}\bigr]}(Z,C)):= -\log\det( K_{\bigl[\begin{smallmatrix}x\\v\end{smallmatrix}\bigr]}(Z,C)K_{\bigl[\begin{smallmatrix}x\\y\end{smallmatrix}\bigr]}^{-1}) + \\ \trace{K_{\bigl[\begin{smallmatrix}x\\y\end{smallmatrix}\bigr]}^{-1}K_{\bigl[\begin{smallmatrix}x\\v\end{smallmatrix}\bigr]}(Z,C)}.
\end{split}
\end{equation} 
Its first variation is provided by (\ref{myeqtop}) at the top of the page.

By the properties of the trace and the Hermitian symmetry, we get that the first variation vanishes if and only if 
\begin{equation}
\trace{\left[\left((K_{xz}K_{zz}^{-1})^*\Delta_{12} + Z^*K_{zz}^{-1}\Delta_{22}\right)\delta Z + C^*\Delta_{22}\delta C \right]} = 0. 
\end{equation}
This holds for all $\delta Z$, $\delta C$ if and only if
\begin{equation}\label{eq:conditions}
\begin{cases}
(K_{xz}K_{zz}^{-1})^*\Delta_{12}+K_{zz}^{-1}Z^*\Delta_{22} = 0\\
C^*\Delta_{22} = 0\\
\end{cases}
\end{equation}
The first equation in \eqref{eq:conditions} can be simplified so that it reads
\begin{equation}\label{eq:cond1_simplified}
K_{xz}\Delta_{12}+Z^*\Delta_{22}=0.
\end{equation}
The matrix inversion lemma allows to compute an explicit expression for matrix $\Delta$
\begin{align*}
\Delta_{12} &= -K_{xx}^{-1}K_{xy}(K_{yy}-K_{yx}K_{xx}^{-1}K_{xy})^{-1} 
+\\&\quad K_{xx}^{-1}K_{xz}K_{zz}^{-1}Z^* \times \\ 
& \left[ZK_{zz}^{-1}(K_{zz}-K_{zx}K_{xx}^{-1}K_{xz})K_{zz}^{-1}Z^* + CC^*\right]^{-1},\\
\Delta_{22} &= \left(K_{yy}-K_{yz}K_{xx}^{-1}K_{xy}\right)^{-1} - \\
& \left[ZK_{zz}^{-1}(K_{zz}-K_{zx}K_{xx}^{-1}K_{xz})K_{zz}^{-1}Z^* + CC^*\right]^{-1}.
\end{align*}
Now, let 
$A := K_{yy}-K_{yx}K_{xx}^{-1}K_{xy}$, and
$B := K_{zz}-K_{zx}K_{xx}^{-1}K_{xz}.$
Then we can write
\begin{align*}
&\Delta_{12} = -K_{xx}^{-1}K_{xy}A^{-1}+ K_{xx}^{-1}K_{xz}K_{zz}^{-1}Z^* \times \\
& \left[ZK_{zz}^{-1}BK_{zz}^{-1}Z^* + CC^*\right]^{-1},\\
&\Delta_{22} = A^{-1} - \left(ZK_{zz}^{-1}BK_{zz}^{-1}Z^* + CC^*\right)^{-1}.
\end{align*}
Therefore, after some manipulation, we conclude that the optimum solution is provided by $C$,$Z$ such that 
\begin{equation}
	\begin{cases}
        C^*= C^*(Z K_{zz}^{-1}B K_{zz}^{-1}Z^* + CC^*)^{-1}A \\
         Z^* = K_{zx}K_{xx}^{-1}K_{xy}+ BK_{zz}^{-1}Z^* \times \\
         \quad (ZK_{zz}^{-1}BK_{zz}^{-1}Z^*+CC^*)^{-1}A
        \end{cases}.
\end{equation}	


\begin{thebibliography}{90}
\bibitem{Bloch-book}
  M. Bloch and J. Barros, {\em Physical-Layer Security. From Information Theory to Security Engineering}. Cambridge University Press, 2011.
\bibitem{Renna} F. Renna, N. Laurenti, H. V. Poor, "Physical-Layer Secrecy for OFDM Transmissions Over Fading Channels," {\em IEEE Trans. Information Forensics and Security}, vol. 7, no. 4, pp. 1354--1367, Aug. 2012.
\bibitem{Tomasin-MIMO} S. Tomasin "Resource allocation for secret transmissions over MIMOME fading channels”, in Proc. {\em IEEE Global Conference on Commun. (GLOBECOM)}, Atlanta, Georgia, Dic. 2013.
\bibitem{Maurer00} U.M. Maurer, ``Authentication theory and hypothesis testing,'' {\it IEEE Trans.\ Inf.\ Theory}, vol.~46, Jul.\ 2000, pp.~1350--1356.
\bibitem{Lai09} L. Lai, H. El Gamal, and H. V. Poor ``Authentication Over Noisy Channels'', {\it IEEE Trans.\ Inf.\ Theory}, vol.~55, no.~2, pp. 906--916, Feb.~2009.
\bibitem{Daniels05} T. Daniels, M. Mina, and S.F. Russell, ``A Signal Fingerprinting Paradigm for General Physical Layer and Sensor Network Security and Assurance,'' {\it IEEE SECURECOMM}, pp.~1-3, Athens (Greece), Sep.\ 2005.
\bibitem{Faria06} D.B. Faria and D.R. Cheriton, ``Detecting identity-based attacks in wireless networks using signalprints,'' {\it ACM WiSe}, pp.~43--52, Los Angeles (CA), Sep.\ 2006.
\bibitem{Xiao09} L. Xiao, L.J. Greenstein, L. Fellow, N.B. Mandayam, and W. Trappe, ``Channel-based spoofing detection in frequency-selective Rayleigh channels,'' {\it IEEE Trans.\ Wireless Commun.}, vol.~8, 2009, pp.~5948--5956.
\bibitem{Baracca12} P. Baracca, N. Laurenti, and S. Tomasin, ``Physical layer authentication over MIMO fading wiretap channels,'' {\it IEEE Trans.\ Wireless Commun.}, vol.~11, 2012, pp.~2564--2573

\bibitem{Cachin98} C. Cachin, ``An Information-Theoretic Model for Steganography,'' in {\it International Workshop on Information Hiding, IH'98}, Portland, OR, April 14--17, 1998, vol. LNCS-1525, pp. 306--318.

\bibitem{Barni13} M. Barni, and B. Tondi, ``The Source Identification Game: An Information-Theoretic Perspective,'' {\it IEEE Trans.\ on Inform.\ Forens.\ Security}, vol. 8, no. 3, pp. 450--463, Mar.~2013.

\bibitem{Kay93} S. M. Kay, {\it Fundamentals of statistical signal processing. Estimation theory}, Prentice Hall, 1993.
\bibitem{Kullback67} S. Kullback, {\it Information Theory and Statistics}, Dover Publications, NY, 1967.
\bibitem{Speed86} T. P. Speed, and H. T. Kiiveri, ``Gaussian Markov Distributions over Finite Graphs'', {\it Annals of Statistics}, vol.~14, no.~1, pp.138--150, Mar.~1986. 
\bibitem{Dempster-72}
A. P. Dempster, ``Covariance selection,''
{\em Biometrics}, vol. 28, 1972, pp.~157--175.
\bibitem{Ferrante-P-IEEE-IT-11}
A.~Ferrante and M.~Pavon, ``Matrix Completion {\em \`a la} Dempster by the Principle of Parsimony,''
{\em IEEE Trans. Information Theory}, vol. 57, 2011, pp.~3925--3931.
  
  
\bibitem{CFPP-IEEE11}
F.~Carli, A.~Ferrante, M.~Pavon, and G.~Picci, 
``A Maximum Entropy Solution of the Covariance
Extension Problem for Reciprocal Processes,''
{\em IEEE Trans. Automatic Control}, vol. 56, 2011, pp.~1999--2012.

\bibitem{Neeser93} F. D. Neeser, and J. L. Massey, ``Proper Complex Random Processes with Applications to Information Theory'', {\it IEEE Trans. \ Inf. \ Theory}, vol.~39, no.~4, pp.1293--1302, Jul.~1993. 

\bibitem{Deuschel-Stroock}
J.-D. Deuschel, and D. W. Stroock, {\em Large deviations.} Academic Press, New York, 1989.













\end{thebibliography}
\end{document}